# Universal kinetics of the stochastic formation of polarization domain structures in a uniaxial single-crystal ferroelectric

Olga Y. Mazur[1,2], Leonid I. Stefanovich[3], and Yuri A. Genenko[2],

[1]Institute of Mechatronics and Computer Engineering, Technical University of Liberec, Studentská 1402/2, 461 17 Liberec 1, Czech Republic

[2]Institute of Materials Science, Technical University of Darmstadt, Otto-Berndt-Str, 3, 64287 Darmstadt, Germany

[3]Branch for Physics of Mining Processes of the M.S. Poliakov Institute of Geotechnical Mechanics of the National Academy of Sciences of Ukraine, 49600 Dnipro, Simferopolska st., 2a, Ukraine

## Abstract

Initial conditions after quenching from a high-temperature paraelectric phase to a low-temperature ferroelectric phase have a substantial impact on the temporal development and formation of stable polarization domain structures which eventually determine physical properties and the functionality of ferroelectrics. Based on the recently advanced exactly solvable model of the stochastic domain structure kinetics in a uniaxial ferroelectric [Phys. Rev. B 107, 144109 (2023)], we study the effect of the magnitude of the initial disorder, its initial correlation length and polarization correlation function on the system evolution. For different shapes of the initial correlation function, the time-dependent correlation length and the two-point polarization correlation coefficient are calculated analytically demonstrating universal features and a good agreement with the available experimental data. Particularly, the magnitude of the charge density correlation function reveals a strong reduction of the bound charges at the nominally charged domain walls which was recently observed experimentally in uniaxial ferroelectrics. Consequently, the integrodifferential equations of evolution for the polarization correlation function and the mean polarization are numerically solved for different initial conditions. The temporal dependence of the polarization mean value and variance are evaluated demonstrating the bifurcation behavior depending on the applied electric field. The impact of the initial state properties on the coercive field deciding between the single- and multi-domain final states of the system is disclosed.

## 1. Introduction

Quenching from a high-temperature to a low-temperature phase has a versatile effect on different physical properties of materials. Particularly, quenching temperature and/or cooling rate have a significant effect on the formation of polarization domain structure and consequent application-related properties of ferroelectrics [1-7]. Thus, in various single crystalline materials the cooling-rate influence on the domain structure evolution was observed [1-6] which allowed control of the domain wall mobility in telluric acid ammonium phosphate (TAAP) [1] and triglycine sulfate (TGS) [2] single crystals, domain shape and size in lead germanate (PGO) single crystals [3] and enhancement of the dielectric constant in lead-magnesium niobate/lead titanate (PMN-PT) single crystals [4,5]. Particularly, in situ 3D observation of the domain wall dynamics in TGS single crystal was observed at different quenching temperatures by second-harmonic generation microscopy [6]. Significant effect of the cooling rate on the formation of domain structure and the consequently enhanced piezoelectric properties was observed on bismuth ferrite–barium titanate (BFO-BT) ferroelectric ceramics [7]. The quenching and cooling rate affected also the phase symmetry, characteristic temperatures and phase composition in sodium bismuth titanate–barium titanate (NBT-BT) relaxor ferroelectrics [8-10] allowing the improvement of their functional properties.

Additional possibilities for controlling the domain structure development are provided by cooling in the external electric or elastic field. Thus, the cooling rate effect on the formation of ferroelastic domain structures in strained ferroelectric films was observed on lead titanate (PT) [11] and lead zirconate titanate (PZT) [12] epitaxial thin films. For three-dimensional samples, the effect of hydrostatic pressure on the phase diagram and domain formation kinetics was studied theoretically in potassium nitrate and potassium dihydrogen phosphate (KDP) crystals [13] and experimentally and theoretically in BT [14]. Alternatively, when applying an electric field, different domain pattern evolutions were observed in PGO [3], TGS [15] and TAAP [16] with and without the field.

Formation of polarization domain structures in ferroelectrics after quenching is a stochastic process and it was previously treated by means of stochastic theories. In the work by Kalkum et al. [17] the formation of periodic domain structures in a uniform system subject to spatially modulated electric field was considered based on the approach developed by Ishibashi and Takagi [18]. Alternatively, Rao and Chakrabarti advanced a random-field model of domain growth based on the stochastically extended Landau-Ginzburg-Devonshire (LGD) approach [19]. In the mentioned models, however, the feedback via electric depolarization fields emerging together with the domain appearance was not accounted. Darinskii et al. developed an LGD-based stochastic approach for the two-component polarization with the self-consistent account of the emerging depolarization fields [20]. In their work, the transient development of the inhomogeneous phase was not studied and correlation functions of the stochastic variables were not considered which is the focus of the current study.

Recently, the authors advanced an exactly solvable model of the stochastic domain formation in uniaxial ferroelectrics which accounts self-consistently for electric interaction between emerging polarization domains [21]. In the current work, this model is used to investigate, to what extend the temporal development of the domain structure depends on and can be controlled by initial conditions and the applied electric field. The paper is organized as follows. In section 2, the basic assumptions of the stochastic model and the underlying Landau-Ginzburg-Devonshire thermodynamic potential are formulated. In section 3, the evolution equations for the stochastic polarization and the electric potential are briefly presented as well as the previously derived system of the integrodifferential equations for the mean polarization and the polarization correlation functions [21]. The solution of these equations is then reduced to the numerical analysis of the system of nonlinear differential equations for the mean value and variance of the polarization. In section 4, analytical expressions for the correlation length and the polarization correlation coefficient are derived for three different shapes of the initial polarization correlation function and compared with the available experimental data for TGS. In section 5, the domain formation kinetics is studied by the numerical solution of the coupled differential equations for the mean value and variance of the polarization for the three cases of the above introduced initial correlation functions. The charge density fluctuations and the implications for the charged domain walls are then studied in section 6. Finally, the physical results obtained in sections 4-6 are discussed in section 7 and concluded in section 8.

## 2. The physical model

Let us consider a uniaxial single-crystalline ferroelectric with the polarization along the $z$-axis of the Cartesian coordinate system as is shown in Fig. 1. It delineates a typical experimental geometry of a ferroelectric plate of thickness $h_f$, attached to a bottom electrode and a dielectric layer of thickness $h_d$ at the top side covered with a top electrode allowing for application of an external voltage. The LGD energy functional of the system can be presented in the form [22,23]

$$\Phi = \Phi_0 + \int_{V_f} \left[\frac{1}{2} A P_z^2 + \frac{1}{4} B P_z^4 + \frac{1}{2} G (\nabla P_z)^2 - P_z E_z - \frac{\varepsilon_0 \varepsilon_b}{2} \boldsymbol{E}^2\right] dV - \int_{V_d} \frac{\varepsilon_0 \varepsilon_d}{2} \boldsymbol{E}^2 dV \quad (1)$$

with the temperature dependent coefficient $A = \alpha_0(T - T_c)$, $\alpha_0 > 0$, $T < T_c$, which is the temperature of the second order paraelectric-ferroelectric phase transition, and the other temperature independent coefficients $B > 0$ and $G > 0$. $\boldsymbol{E}$ denotes the local electric field, $\varepsilon_0$, $\varepsilon_d$ and $\varepsilon_b$ are the permittivity of vacuum, of the dielectric layer and the background permittivity of the ferroelectric, respectively, while $V_f$ and $V_d$ denote the volumes of the ferroelectric plate and the dielectric layer, respectively.

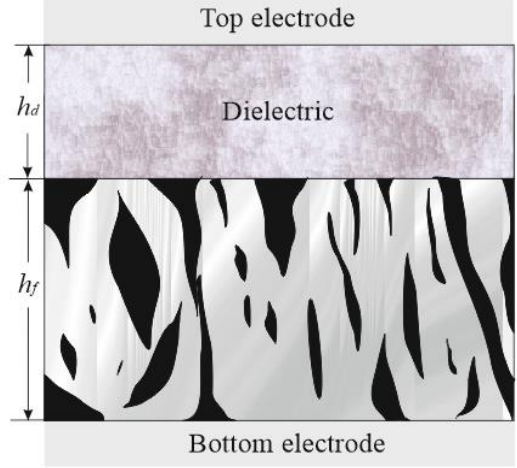

Fig. 1. A ferroelectric single crystal of thickness $h_f$ placed on a bottom electrode and separated from a top electrode by a vacuum or dielectric layer of thickness $h_d$ is infinite in $(x, y)$ plane parallel to the ferroelectric surface. The only polarization direction is along the vertical $z$-axis of the Cartesian $(x, y, z)$-frame.

We use in the following dimensionless physical variables normalized to their natural characteristic values in the phase transition problem. Thus, we denote a dimensionless polarization $\pi = P_z/P_s$ normalized to the spontaneous equilibrium polarization $P_s = \sqrt{|A|/B}$, and a dimensionless electric field $\epsilon = E/E_0$ with the value of $E_0 = P_s|A|$. Spatial coordinates are normalized to a characteristic length $\lambda = \sqrt{G/|A|}$, which has a meaning of the characteristic domain wall thickness, and the time $t$ is normalized to $t_0 = \Gamma/|A|$ as $\tau = t/t_0$ with the Khalatnikov constant $\Gamma$ [21].

We consider evolution of the system from an initial state obtained by quenching from a high temperature paraelectric phase to a ferroelectric one at temperature $T < T_c$. Since the initial state is a random one, all physical variables become stochastic variables too and will be treated in this model as Gauss random fields as suggested previously [24,25]. If temperature $T$ is not too close to $T_c$ and the magnitude and spatial scale of disorder are large enough, the development of the system is dominated by the quenched disorder while the thermal fluctuations can be neglected as was shown in Ref. [21].

The polarization can be represented as $\pi(\mathbf{r}, \tau) = \bar{\pi}(\tau) + \xi(\mathbf{r}, \tau)$ with the dimensionless mean polarization magnitude $\bar{\pi}(\tau) = \langle P_z(\mathbf{r}, \tau) \rangle / P_s$, depending on the dimensionless time $\tau$, and the stochastic polarization $\xi(\mathbf{r},\tau)$ such that $\langle\xi(\mathbf{r},\tau)\rangle=0$. Here the sign $\langle...\rangle$ denotes statistical averaging over all possible system realizations. Then a dimensionless local electric field reads

$$\boldsymbol{\epsilon}(\mathbf{r},\tau)=\epsilon_a\hat{\mathbf{z}}-\alpha_z\bar{\pi}(\tau)\hat{\mathbf{z}}-\boldsymbol{\nabla}\phi(\mathbf{r},\tau). \tag{2}$$

Here a uniform electric field $\epsilon_a$ in the ferroelectric in z-direction is determined by a voltage $V$ applied to the electrodes and the chosen sample geometry and material parameters [21],

$$\epsilon_a=\frac{\varepsilon_d V}{(\varepsilon_d h_f+\varepsilon_b h_d)E_0}. \tag{3}$$

The second term in Eq. (2) represents the mean depolarization field in the ferroelectric due to the average polarization $\bar{\pi}$ with the depolarization coefficient $\alpha_z=h_d/[(\varepsilon_d h_f+\varepsilon_b h_d)\varepsilon_0|A|]$, while the last term the stochastic electric depolarization field due to the stochastic electric potential $\phi(\mathbf{r},\tau)$, so that $\langle\boldsymbol{\nabla}\phi\rangle=0$.

### 3. The governing equations

By variation of the energy functional (1) with respect to the polarization and the electric potential, respectively, a closed system of evolution equations for these two variables can be derived [21],

$$\begin{cases}\frac{\partial\pi}{\partial\tau}=\Delta\pi+\pi-\pi^3+\epsilon_z & (4a)\\ \Delta\phi=\eta\frac{\partial\pi}{\partial z} & (4b)\end{cases}$$

where the first one is the time-dependent Ginzburg-Landau (TDGL) equation and the second one is the Poisson equation with a dimensionless susceptibility parameter $\eta=1/(\varepsilon_0\varepsilon_b|A|)$.

The Gauss random variables can be completely characterized by a full set of two-point correlations functions between all variables [26]. It turns out that only two of them are relevant for the temporal development of the considered system from the initial quenched state, namely, the autocorrelation function for the stochastic polarization, $K(\mathbf{s},\tau)=\langle\xi(\mathbf{r}_1,\tau)\xi(\mathbf{r}_2,\tau)\rangle$, with $\mathbf{s}=\mathbf{r}_1-\mathbf{r}_2$, and the cross-correlation function between z-components of polarization and electric field, $\Psi_{zz}(\mathbf{s},\tau)=\langle\epsilon_z(\mathbf{r}_1,\tau)\xi(\mathbf{r}_2,\tau)\rangle$ [21]. All the correlation functions are interrelated with each other as was shown in Refs. [21,27], particularly, $\Psi_{zz}(\mathbf{s},\tau)$ is related to $K(\mathbf{s},\tau)$ as

$$\Delta\Psi_{zz}(\mathbf{s},\tau)=-\eta\frac{\partial^2}{\partial s_z^2}K(\mathbf{s},\tau). \tag{5}$$

In terms of the Fourier transforms defined as

$$K(\mathbf{s},\tau)=\frac{1}{(2\pi)^3}\int d^3q\exp(i\mathbf{q}\mathbf{s})\widetilde{K}(\mathbf{q},\tau), \tag{6a}$$

$$\widetilde{K}(\mathbf{q},\tau) = \int d^3s \exp(-i\mathbf{q}\mathbf{s})K(\mathbf{s},\tau) \tag{6b}$$

this relation can be converted into an explicit algebraic expression,

$$\widetilde{\Psi}_{zz}(\mathbf{q},\tau) = -\eta \frac{q_z^2}{q^2}\widetilde{K}(\mathbf{q},\tau) \tag{7}$$

Thus, the problem of the time-dependent spatial correlations is reduced to the finding of the function $\widetilde{K}(\mathbf{q},\tau)$ and the mean polarization value $\bar{\pi}(\tau)$, for which the system of nonlinear integrodifferential equations was derived previously [21],

$$\begin{cases} \frac{d\bar{\pi}}{d\tau} = \bar{\pi}(1 - \alpha_z - 3K(0,\tau)) - \bar{\pi}^3 + \epsilon_a & (8a) \\ \frac{d\widetilde{K}}{d\tau} = 2\left[1 - 3\bar{\pi}^2 - 3K(0,\tau) - \left(q^2 + \eta\frac{q_z^2}{q^2}\right)\right]\widetilde{K} & (8b) \end{cases}$$

The knowledge of $\widetilde{K}(\mathbf{q},\tau)$ allows evaluation of an important characteristic of the system evolution, the correlation length $L(\tau)$, defined by the relation [24]

$$L^{-2}(\tau) = \int d^3q\, q^2 \widetilde{K}(\mathbf{q},\tau) / \int d^3q\, \widetilde{K}(\mathbf{q},\tau). \tag{9}$$

Another important system characteristic is the polarization dispersion, or variance, $D(\tau) = K(0,\tau)$, which can be expressed as the integral of the Fourier transform (6a).

The first order differential equation Eq. (8b) can be solved allowing the expression of the function $\widetilde{K}(\mathbf{q},\tau)$ through its initial value $\widetilde{K}(\mathbf{q},0)$ defined by correlations in the initial state after quenching [21],

$$\widetilde{K}(\mathbf{q},\tau) = \widetilde{K}(\mathbf{q},0)\mu(\tau)\exp\left[-2\left(q^2 + \eta\frac{q_z^2}{q^2}\right)\tau\right] \tag{10}$$

where the auxiliary function $\mu(\tau)$ is defined as

$$\mu(\tau) = \exp\{2\tau - 6\int_0^\tau d\tau'\,[\bar{\pi}^2(\tau') + D(\tau')]\}. \tag{11}$$

The time behavior of both functions $\mu(\tau)$ and $\widetilde{K}(\mathbf{q},\tau)$ is essentially determined by the variance $D(\tau)$. This makes finding of $D(\tau)$ to the central problem in the study of correlations. The variance can be found by solving the system of nonlinear differential equations [21],

$$\begin{cases} \frac{d\bar{\pi}}{d\tau} = \bar{\pi}(1 - \alpha_z - 3D) - \bar{\pi}^3 + \epsilon_a & (12a) \\ \frac{dD}{d\tau} = [2 - 6\bar{\pi}^2 - 6D - \nu(\tau)]D & (12b) \end{cases}$$

with an auxiliary function

$$\nu(\tau) = \frac{2}{L^2(\tau)} + \frac{1}{2\tau}\left[1 - \frac{2}{\sqrt{\pi}}\frac{\sqrt{2\eta\tau}\exp(-2\eta\tau)}{\mathrm{erf}(\sqrt{2\eta\tau})}\right] \tag{13}$$

In the following, this system of equations will be analyzed and solved numerically for different initial physical conditions, which may be realized in the course of quenching, and different applied field magnitudes. Some characteristics of the system, however, may be found analytically in a closed form as will be shown in the next section.

## 4. Correlation length and correlation coefficients for different initial conditions

We start with a derivation of some useful general relations. When considering only isotropic initial disordered states after quenching, meaning that $\widetilde{K}(\mathbf{q}, 0) = \widetilde{K}(q, 0)$, as it will be done in the following, the functions $L(\tau)$ and $\mu(\tau)$ can be conveniently expressed in terms of an auxiliary function

$$\kappa(\tau) = \int_0^\infty dq\, q^2 \exp(-2\tau q^2)\widetilde{K}(\mathbf{q}, 0) \tag{14}$$

Then, from the definition of $L(\tau)$, Eq. (9), one finds

$$L^{-2}(\tau) = -\frac{1}{2}\frac{\partial}{\partial \tau}\ln \kappa(\tau), \tag{15}$$

which can be directly evaluated from the characteristics of the initial state only. This demonstrates that properties of the initial state are decisive for the subsequent evolution of the system.

Now, by the integration of Eq. (12b) and using the definition of $\mu(\tau)$, Eq. (11), it follows that

$$\mu(\tau) = \frac{D(\tau)}{D(0)}\exp\left(\int_0^\tau d\tau'\, \nu(\tau')\right). \tag{16}$$

By substituting the expression for $\nu(\tau)$, Eq. (13), into Eq. (16) and using the relation (15), Eq. (16) can be transformed to

$$\mu(\tau) = \frac{2}{\sqrt{\pi}}\frac{\sqrt{2\eta\tau}}{\mathrm{erf}(\sqrt{2\eta\tau})}\frac{\kappa(0)}{\kappa(\tau)}\frac{D(\tau)}{D(0)}. \tag{17}$$

Using the definition of $\kappa(\tau)$, Eq. (14), it is easy to check that $\kappa(0) = 2\pi^2 D(0)$.

The properties of the initial quenched state can in principle be different and depend on various factors like quenching temperature, cooling rate, etc., as was discussed in Section 1. We will consider in the following three examples of the initial correlation function $K(\mathbf{s}, 0)$ and study the effect of their shape on the system evolution. One of the reasons of this comparative analysis is that the experimentally observed correlation coefficient $C(s, \tau) = K(\mathbf{s}, \tau)/D(\tau)$ exhibits at small distances linear dependence on $s$ at any times $\tau$ which could not yet be adequately described [21].

### 4.1. Initial correlation function of a Gaussian shape

This case was treated in Ref. [21]. For completeness we shortly reproduce here the main results. Assuming

$$K(s, 0) = K_0 \exp\left(-\frac{s^2}{2r_c^2}\right) \tag{18}$$

with the Gauss parameter $r_c$ and the magnitude $K_0 = D(0)$ we get its Fourier transform

$$\widetilde{K}(\mathbf{q}, 0) = (2\pi)^{\frac{3}{2}} K_0 r_c^3 \exp\left(-\frac{r_c^2 q^2}{2}\right) \tag{19}$$

This choice defines the auxiliary functions

$$\kappa(\tau) = 2\pi^2 K_0 \left(1 + \frac{4\tau}{r_c^2}\right)^{-3/2} \tag{20}$$

and

$$\mu(\tau) = \frac{2}{\sqrt{\pi}} \frac{\sqrt{2\eta\tau}}{\text{erf}(\sqrt{2\eta\tau})} \left(1 + \frac{4\tau}{r_c^2}\right)^{3/2} \frac{D(\tau)}{D(0)}, \tag{21}$$

as well as the correlation length

$$L(\tau) = \sqrt{(r_c^2 + 4\tau)/3}. \tag{22}$$

The according rather formidable expressions for the longitudinal and transverse polarization correlation coefficients may be found in Ref. [21]. The correlation coefficient $C(s, \tau)$ does not exhibit a linear behavior in $s$ for small distances at any time $\tau$ and thus cannot adequately describe experimental observations on TGS presented in many works [15,28-32].

### 4.2. Initial correlation function of an exponential shape

To be able to describe the linear spatial behavior of $C(s, \tau)$ for small distances $s$ at least at the beginning of the system evolution we assume now the initial correlation function of the form

$$K(s, 0) = K_0 \exp\left(-\frac{s}{\xi}\right) \tag{23}$$

with a characteristic length $\xi$. This function has the Fourier transform

$$\widetilde{K}(\mathbf{q}, 0) = \frac{8\pi K_0}{\xi (q^2 + \xi^{-2})^2}. \tag{24}$$

The corresponding auxiliary functions read

$$\kappa(\tau) = 2\pi^2 K_0 \left[(1 + 2\sigma^2) \exp(\sigma^2)\text{erfc}(\sigma) - \frac{2}{\sqrt{\pi}}\sigma\right] \quad (25)$$

and

$$\mu(\tau) = \frac{2}{\sqrt{\pi}} \frac{\sqrt{2\eta\tau}}{\text{erf}(\sqrt{2\eta\tau})} \left[(1 + 2\sigma^2)\exp(\sigma^2)\text{erfc}(\sigma) - \frac{2}{\sqrt{\pi}}\sigma\right]^{-1} \frac{D(\tau)}{D(0)} \quad (26)$$

where a combined variable $\sigma = \sqrt{2\tau}/\xi$ was introduced for convenience.

The correlation length follows then from Eqs. (15) and (25) as

$$L(\tau) = \frac{\xi}{\sqrt{2}} \sqrt{\frac{\sqrt{\pi}\sigma(1 + 2\sigma^2)\exp(\sigma^2)\text{erfc}(\sigma) - 2\sigma^2}{(1 + \sigma^2) - \frac{3}{2}\sqrt{\pi}\,\sigma\left(1 + \frac{2}{3}\sigma^2\right)\exp(\sigma^2)\text{erfc}(\sigma)}} \quad (27)$$

In the limiting case $\tau \to 0$ the characteristic length behaves as $L(\tau) \cong (\pi\xi^2\tau/2)^{1/4}$ and, thus, vanishes at $\tau = 0$, differently from the Gaussian function case, Eq. (22). However, asymptotically, at $\tau \to \infty$, $L(\tau) \sim \sqrt{4\tau/3}$ as in Eq. (22). A comparison of the analytical results (22) and (27) with the experiment by Tomita et al. [15] is shown in Fig. 2.

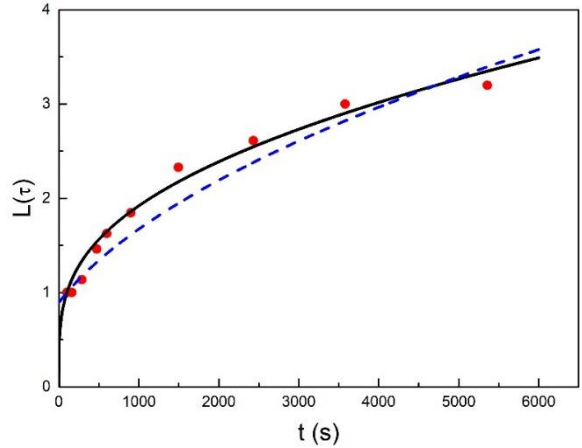

Fig. 2. Fitting the experimental data for $L(\tau)$ from Tomita et al. [15] by Eq. (27) using the fitting parameter $\xi = 1$ and the time scaling $\tau = t/t_0$ with $t_0 = 6000\,s$ (the solid line). The points present experimental results while the dashed line fitting by Eq. (22) using the parameters $r_c = 1$ and $\tau = t/t_0$ with $t_0 = 1600\,s$.

The relevant for experiments transverse correlation coefficient $C(\mathbf{s}, \tau) = K(\mathbf{s}, \tau)/D(\tau)$ for the distance $\mathbf{s} = (\mathbf{s}_\perp, 0)$ can be calculated by substituting Eqs. (24) and (26) into Eq. (10) resulting in the expression

$$C_\perp(\mathbf{s}_\perp,\tau) = \frac{8}{\pi} \frac{\sqrt{2\eta\tau}\exp(-2\eta\tau)}{\mathrm{erf}(\sqrt{2\eta\tau})\left[\sqrt{\pi}(1+2\sigma^2)\exp(\sigma^2)\mathrm{erfc}(\sigma) - 2\sigma\right]} \int_0^\infty dq \frac{q^2 \exp(-\sigma^2 q^2)}{(1+q^2)^2}$$
$$\times \int_0^1 dz \frac{z\exp(2\eta\tau z^2)}{\sqrt{1-z^2}} J_0\left(\frac{qzs_\perp}{\xi}\right). \qquad (28)$$

As is seen from the example evaluations of Eq. (28) for different times $\tau$ in Fig. 3, the function $C(\mathbf{s},\tau)$ exhibits initially a linear dependence on $s_\perp$ but then quickly transforms with increasing time to a curve with a smooth maximum.

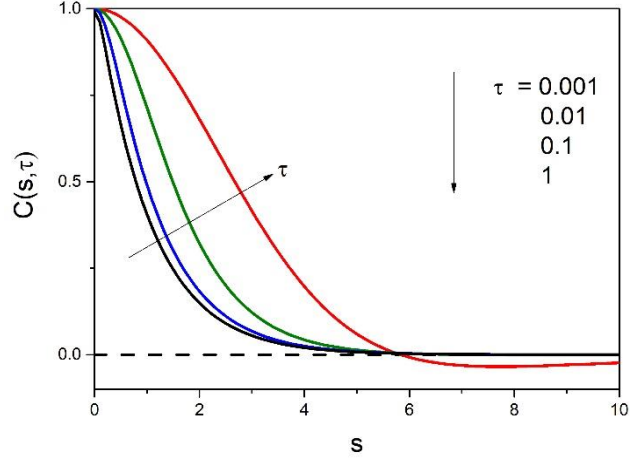

Fig. 3. The time development of the spatial dependence of the correlation coefficient $C_\perp(\mathbf{s}_\perp,\tau)$ using the parameter $\xi = 1$ and $\eta = 1$, and represented for the dimensionless times $\tau = 0.001, 0.01, 0.1$ and $1$.

A comparison of Eq. (28) with the experimental data for the correlation coefficient $C_\perp(\mathbf{s}_\perp,\tau)$ from Tomita et al. [15] is shown in Fig. 4. Hereby the parameters $\xi = 1$ and $t_0 = 6000\,s$ are taken over from the fitting of $L(\tau)$ in Fig. 2 while the dimensionless polarizability $\eta = 10$ is chosen by the best fit of the experimental data on correlations.

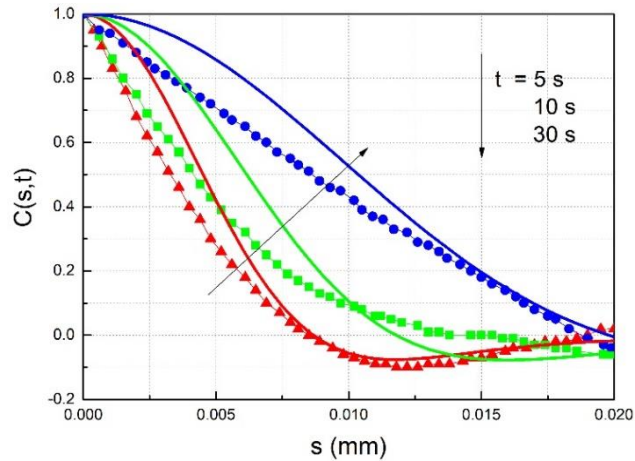

Fig. 4. The correlation coefficient $C_\perp(\mathbf{s}_\perp,\tau)$ evaluated using Eq. (28) for different times as is shown in the plot (solid lines) is compared with the experimental data by Tomita et al. [15]

represented for the respective times by solid symbols. The parameters $\xi$ and $t_0$ in Eq. (28) are taken over from the fitting of $L(\tau)$ in Fig. (2) while the dimensionless polarizability $\eta = 10$ was chosen by the best fit of the experimental correlation data.

While the global spatial dependence is in principle captured by the formula (28), including the minimum in the negative region, the linear spatial dependence at the origin is missing in the considered time region since the spatial dependence of $C_\perp(\mathbf{s}_\perp, \tau)$ transforms quickly from a sharp to a smooth maximum as demonstrated in Fig. 3. Thus, the assumption of the initially linear correlation function, Eq. (23), does not help to improve the spatial dependence of $C_\perp(\mathbf{s}_\perp, \tau)$ at later times.

### 4.3. Initial correlation function of an error function type

Let us consider another case with an initial linear spatial dependence at small distances given by the complementary error function

$$K(\mathbf{s}, 0) = K_0 \operatorname{erfc}\left(\frac{s}{\xi}\right) \tag{29}$$

with the Fourier transform

$$\widetilde{K}(\mathbf{q}, 0) = -\frac{4\sqrt{\pi} K_0 \xi}{q^2}\left[1 + \frac{i\sqrt{\pi}}{q\xi}\left(1 + \frac{q^2\xi^2}{2}\right)\exp\left(-\frac{q^2\xi^2}{4}\right)\operatorname{erf}\left(\frac{iq\xi}{2}\right)\right]. \tag{30}$$

This results in the auxiliary functions

$$\kappa(\tau) = 4\pi K_0\left[\arctan\left(\frac{1}{2\sigma}\right) - \frac{\sigma}{2(\sigma^2 + 1/4)}\right] \tag{31}$$

with the combined variable $\sigma = \sqrt{2\tau}/\xi$ as above and

$$\mu(\tau) = \frac{\sqrt{2\pi\eta\tau}}{\operatorname{erf}(\sqrt{2\eta\tau})}\left[\arctan\left(\frac{1}{2\sigma}\right) - \frac{\sigma}{2(\sigma^2+1/4)}\right]^{-1}\frac{D(\tau)}{D(0)}. \tag{32}$$

The correlation length follows then from Eqs. (15) and (31) as

$$L(\tau) = 2\xi(\sigma^2 + 1/4)\sqrt{2\sigma \arctan\left(\frac{1}{2\sigma}\right) - \frac{\sigma^2}{\sigma^2+1/4}}. \tag{33}$$

In the limiting case $\tau \to 0$ the characteristic length behaves as $L(\tau) \cong (\pi^2\xi^2\tau/8)^{1/4}$ and, thus, vanishes at $\tau = 0$, differently from the Gaussian function case, Eq. (22). However, asymptotically, at $\tau \to \infty$, again $L(\tau) \sim \sqrt{4\tau/3}$ as in Eq. (22). A comparison of the analytical results (22) and (33) with the experiment by Tomita et al. [15] is shown in Fig. 5.

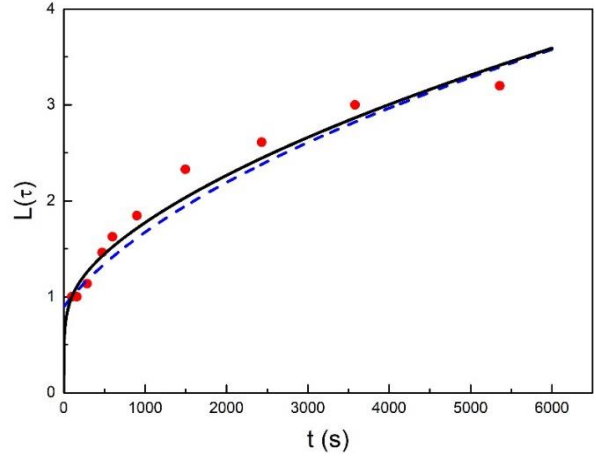

Fig. 5. Fitting the experimental data for $L(\tau)$ from Tomita et al. [15] by Eq. (33) using the fitting parameter $\xi = 1$ and the time scaling $\tau = t/t_0$ with $t_0 = 6400\ s$ (the solid line). The points represent experimental results and the dashed line fitting with Eq. (22) using the parameters $r_c = 1$ and $\tau = t/t_0$ with $t_0 = 1600\ s$.

The experimentally measurable transverse correlation coefficient $C(\mathbf{s}, \tau) = K(\mathbf{s}, \tau)/D(\tau)$ for the distance $\mathbf{s} = (\mathbf{s}_\perp, 0)$ can be calculated by substituting Eqs. (30) and (32) into Eq. (10) resulting in the expression

$$C_\perp(\mathbf{s}_\perp, \tau) = \frac{-2}{\pi} \frac{\sqrt{2\eta\tau}\exp(-2\eta\tau)}{\mathrm{erf}(\sqrt{2\eta\tau})\left[\arctan\left(\frac{1}{2\sigma}\right) - \frac{\sigma}{2(\sigma^2 + 1/4)}\right]} \int_0^1 dz \frac{z\exp(2\eta\tau z^2)}{\sqrt{1-z^2}}$$

$$\times \int_0^\infty dq\, \exp(-\sigma^2 q^2) J_0(qzs_\perp/\xi) \left[1 + \frac{i\sqrt{\pi}}{q}\left(1 + \frac{q^2}{2}\right)\exp\left(-\frac{q^2}{4}\right)\mathrm{erf}\left(\frac{iq}{2}\right)\right] \quad (34)$$

As well as in the case of the initial exponential correlation function, the example evaluation of Eq. (34) for different times $\tau$ in Fig. 6 exhibits initially a linear dependence on $s_\perp$ which then quickly transforms with increasing time to a curve with a smooth maximum.

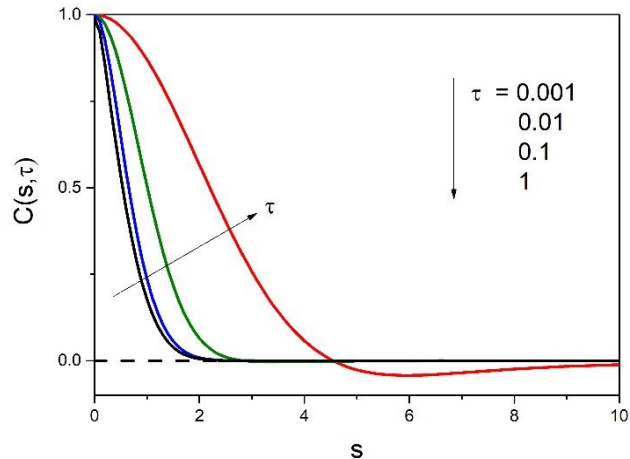

Fig. 6. The time development of the spatial dependence of the correlation coefficient $C_\perp(\mathbf{s}_\perp, \tau)$ using the parameter $\xi = 1$ and $\eta = 1$, and represented for the dimensionless times $\tau = 0.001, 0.01, 0.1$ and $1$.

A comparison of Eq. (34) with experimental data for the correlation coefficient $C_\perp(\mathbf{s}_\perp, \tau)$ from Tomita et al. [15] shown in Fig. 7 demonstrates the traits similar to the case of the initial exponential correlation function, Fig. 4. For the evaluation of Eq. (34), the parameters $\xi = 1$ and $t_0 = 6400\ s$ were taken over from the fitting of $L(\tau)$ in Fig. 5 while the dimensionless polarizability $\eta = 10$ was chosen by the best fit of the experimental correlation data.

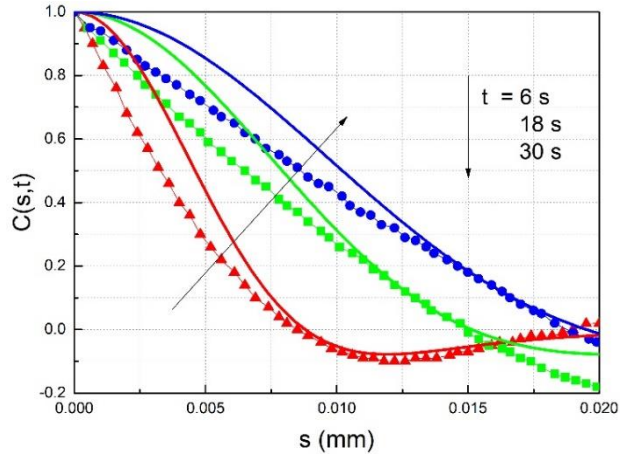

Fig. 7. The correlation coefficient $C_\perp(\mathbf{s}_\perp, \tau)$ evaluated using Eq. (34) for different times as is shown in the plot (solid lines) is compared with the experimental data by Tomita et al. [15] represented for the respective times by solid symbols. The parameters $\xi$ and $t_0$ in Eq. (34) are taken over from the fitting of $L(\tau)$ in Fig. (5).

Again, the global spatial dependence provided by the formula (34) captures, in principle, the experimental behavior including the minimum in the negative region, but the linear spatial dependence at the origin is missing in the considered time region since the spatial dependence of $C_\perp(\mathbf{s}_\perp, \tau)$ quickly transforms with time to a smooth maximum as can be expected considering Fig. 6. Thus, the assumption of the initially linear correlation function of any shape, exemplified above by Eqs. (23) and (29), does not help to improve the spatial dependence of $C_\perp(\mathbf{s}_\perp, \tau)$ at later times. This suggests that the spatial behavior of the correlation coefficient $C_\perp(\mathbf{s}_\perp, \tau)$ is rather determined by the structure of Eq. (8b) than by the initial condition to this equation, $\widetilde{K}(\mathbf{q}, 0)$. Furthermore, the apparent discrepancy between the experimental data and the theory near the origin may have another reason which will be discussed in the next section.

### 4.4. Spatial dependence of model correlation functions

Following the pioneering works by the Ishibashi group [15,28] the statistical treatment of the stochastic polarization development in TGS was performed by other groups [29-32] in the same spirit of the Ising model, namely, by assigning the values of the scalar order parameter +1 and -1 inside ''black'' and ''white'' domains, respectively, reflecting the two opposite polarization states. This simplified treatment excludes a possible smooth spatial variation of the polarization and, thus, assumes the domain walls of zero thickness. To comprehend, what role plays this art of treatment of experimental data we will consider below a couple of insightful examples of correlation functions.

To this end we take two periodic stripe domain structures of zero total polarization with zero and non-zero domain wall thickness, respectively, delineated in Fig. 8(a). The smooth polarization distribution represented by the dashed line in Fig. 8(a) is described by a harmonic function,

$$\pi(x) = \cos(\pi x/2h), \qquad (35)$$

with the parameter $h = 1/4$, while the step-like polarization stripe structure consists of positive and negative domains of the equal widths $a = b = 1/2$, respectively.

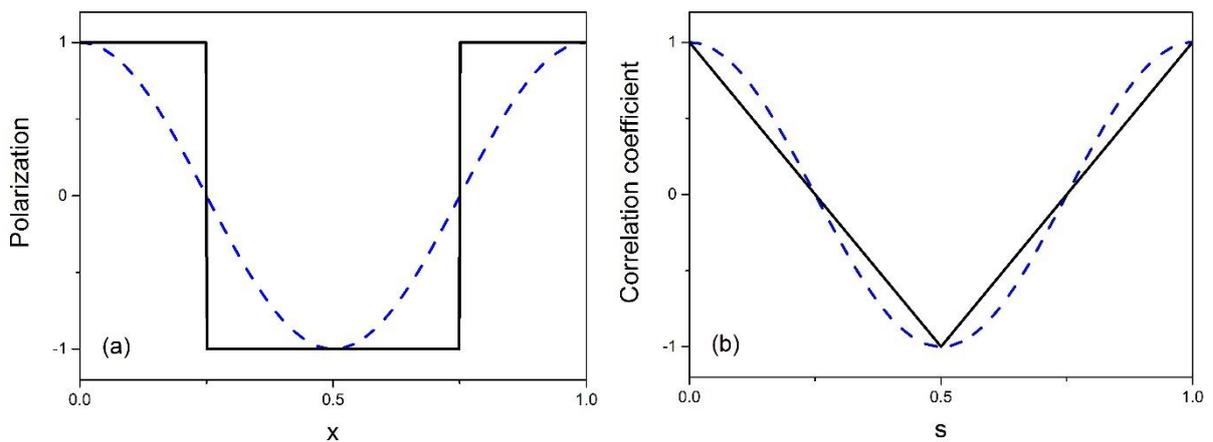

Fig. 8. (a) Periodic normalized polarization profiles with zero (solid line) and non-zero (dashed line) domain wall thicknesses. Both polarization stripe structures have a period of unity with positive and negative polarization domains of the width ½. (b) Correlation coefficients for the polarization distributions in (a) shown respectively by the solid and dashed lines.

Since these structures exhibit long-range order, they are, strictly speaking, not stochastic. Nevertheless, they can be characterized by a spatially averaged two-point correlation function in the same way as it is done in experiments [15, 29-32]. Using the results of Appendix B of Ref. [21], the polarization correlation coefficients for the two polarization distributions depicted in Fig. 8(a) can be represented by plots in Fig. 8(b).

The smooth correlation coefficient represented by the dashed line in Fig. 8(b) corresponds to the harmonic polarization distribution shown also by the dashed line in Fig. 8(a). It is described, for the chosen parameter $h = 1/4$, by the function $C(s) = \cos(2\pi s)$. The correlation coefficient, represented piecewise by straight solid lines in Fig. 8(b), corresponds to the step-like polarization distribution shown by solid lines in Fig. 8(a). Starting from the origin, this correlation coefficient is described by the dependence $C(s) = 1 - 4s$. When applying the definition of the characteristic length $L$ by the condition $C(s) = 1/2$, as in the experiments [15, 29-32], one obtains $L = 1/8$ for the latter function. Using this, the correlation coefficient may be expressed as $C(s) = 1 - s/2L$ coinciding with the scaling behavior near the origin declared in the experiments [15, 29-32]. Notably, this behavior is observed for the structure with the abrupt polarization variation and the zero-thickness domain walls, while the smeared polarization distribution in Fig. 8(a) is characterized by the correlation coefficient with a smooth maximum in Fig. 8(b). This suggests, that the apparent linear spatial behavior of the experimental correlation coefficients might originate from the way of the experimental data treatment neglecting spatially smeared polarization variations.

## 5. Effect of initial conditions on the domain formation kinetics and the coercive field

In this section, we study the effect of different initial shapes of the polarization correlation function as well as the effect of the magnitude of the initial spatial scale and the initial disorder on the temporal domain structure formation at different applied electric field values.

Since the polarization correlation function $K(\mathbf{s}, \tau)$ is completely determined by its initial shape $K(\mathbf{s}, 0)$ via Eq. (10) and the auxiliary time-dependent function (17), the time development of the system is fully governed by the variance $D(\tau)$ and the average polarization $\bar{\pi}(\tau)$. The latter two dependences may be derived by solving the system of nonlinear differential equations (12). This is performed here numerically by means of the Matlab package (MATLAB ver. R2023b). Based on the above fitting of the experiments on TGS by Tomita et al. [15] in Figs. 4 and 7 we fix the dimensionless parameter of susceptibility at $\eta = 10$. The geometrical dimensions of the structure in Fig. 1 necessary for calculations are chosen as $h_f = 1$ mm and $h_d = 20$ μm which is typical for experiments [15, 29-32]. The dielectric permittivity of the materials is taken as $\varepsilon_b = 10$ and $\varepsilon_d = 3$. This choice leads to the value of the parameter $\alpha_z = h_d \varepsilon_b \eta/(\varepsilon_d h_f + \varepsilon_b h_d) = 0.625$. The other involved parameters are varied.

### 5.1. Development from the initial correlation function of a Gaussian shape

In order to comprehend typical behavior patterns of the variables $\bar{\pi}(\tau)$ and $D(\tau)$ we first consider example solutions of the equation system (12) for the initial correlation function of the Gaussian shape, Eq. (18), fixing the initial values at $\bar{\pi}(0) = 0, D(0) = 0.1, r_c = 1$ and varying the applied field values. The choice of the initial correlation function $K(\mathbf{s},0)$ is reflected in Eq. (13) via the correlation length $L(\tau)$ which is taken in this case in the form of Eq. (22). The results are presented on the phase diagram in coordinates $(\bar{\pi},D)$ for different values of the field $\epsilon_a$ as indicated in Fig. 9.

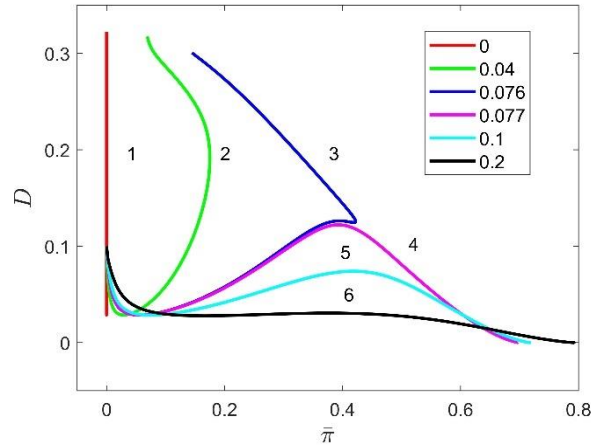

Fig. 9. Phase trajectories of the system starting from the initial values $\bar{\pi}(0) = 0$ and $D(0) = 0.1$ for the parameter values $r_c = 1, \eta = 10, \alpha_z = 0.625$. Curves 1–6 correspond to the values of the dimensionless external electric field in the sample $\epsilon_a = \{0; 0.04; 0.076; 0.077; 0.1; 0.2\}$, respectively.

Development of domain structure after quenching is occurring nonmonotonically and differs depending on the value of applied electric field. The pattern of trajectories in Fig. 9 exhibits bifurcation behavior near the coercive field value about $\epsilon_{cr} = 0.0765$ separating the regions of polydomain and single-domain ordering. In the absence of the external electric field, domains develop nonmonotonically keeping $\bar{\pi} = 0$ while the dispersion $D$ first decreases and then increases to finally generate a quasi-periodic structure with the equal volumes of domains of different signs (curve 1). When the electric field is imposed on the sample, the fraction of domains of the corresponding direction is increased both during the relaxation and at its completion (curve 2).

The most drastic deviations are clearly visible when fields close to the coercive value are applied to the sample (curves 3, 4 in Fig. 9). When $\epsilon_a \simeq \epsilon_{cr}$, many domains appear directed along the field, which is manifested by a weaker increase of the variance $D$ compared to the average polarization $\bar{\pi}$. A pronounced asymmetry of the domains along and opposite to the field is observed in the bifurcation region, and the system exhibits slowing down for some time which can be evaluated as the length of the plateau on the time dependences $\bar{\pi}(\tau)$ and $D(\tau)$ in Fig. 10.

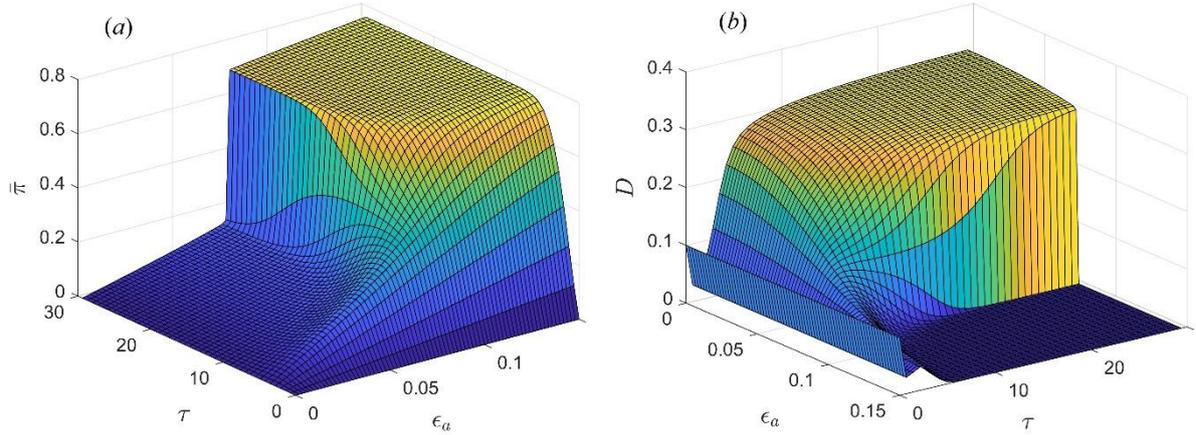

Fig. 10. (a) The time development of the mean polarization for different magnitudes of the electric field. (b) The time development of the polarization variance for different magnitudes of the electric field. All other parameters are the same as used in Fig. 9.

The closer is the applied field $\epsilon_a$ to the coercive field $\epsilon_{cr}$ the more extended in time is the "indefinite" state of the system deciding between the single-domain and multi-domain final state (the so called "kinetic slowing down"). After this interim phase the system tends to balance the ratio of oppositely directed domains when $\epsilon_a \lesssim \epsilon_{cr}$ (curve 3 in Fig. 9) or to polarize the entire sample in one direction when $\epsilon_a \gtrsim \epsilon_{cr}$ (curve 4). The average polarization of the formed polydomain structure is $\bar{\pi}(\tau) > 0$ that still indicates some imbalance of domain fractions (curve 3). A further increase in the magnitude of the applied field leads to the fast formation of a single-domain state (curve 6).

The nonmonotonic behavior of the evolution curves is also manifested by a decrease in the variance $D$ at the beginning of relaxation when the electric field is imposed on the sample. It might indicate both the presence of an incubation period for a nuclei of the ferroelectric domain (curves 2–5) and the suppression of the initial inhomogeneity by the large external electric field (curve 6). Starting from the state with $D(0) = 0.1$ all the curves go through an incubation stage with the same lower magnitude of $D$, as is seen in Fig. 9 and Fig. 10(b), and then develop differently depending on the applied field. It is important to note that the existence of such an incubation period is determined solely by the prehistory of the sample, and is observed even in the absence of an external electric field (Fig. 10(b)).

The magnitude of the initial disorder $D(0)$, which can be experimentally controlled, for example, by the cooling rate, plays a significant role for the system evolution. Since the current model, assuming the dominance of the quenched disorder over the thermal one, applies for the substantial disorder only [21], we consider here the values $D(0) > 0.1$. The phase diagram on the plane ($\bar{\pi}$,D) in Fig. 11 demonstrates the evolution of the system starting from different magnitudes of the initial disorder within the range $0.1 < D(0) < 0.9$ when the applied field $\epsilon_a = 0.09$ is fixed. It is seen that the level of disorder $D$ in the incubation stage, to which the system falls after the start, increases monotonically with $D(0)$.

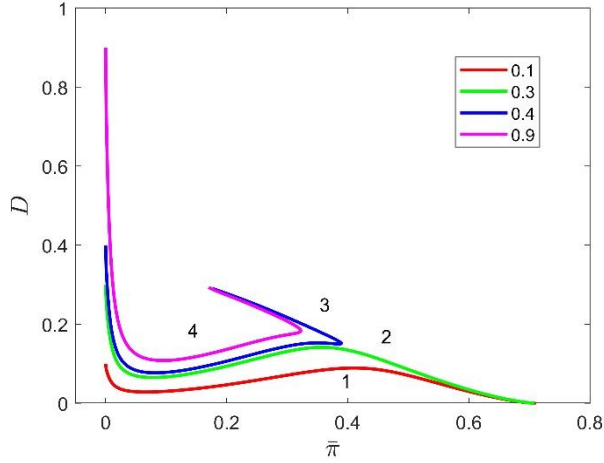

Fig. 11. Phase trajectories of the system starting from the initial value $\bar{\pi}(0) = 0$ under the applied field of $\epsilon_a = 0.09$ for the parameter values $r_c = 1, \eta = 10, \alpha_z = 0.625$. Curves 1–6 correspond to the values of the initial disorder $D(0) = \{0.1; 0.3; 0.4; 0.9\}$, respectively.

The coercive field $\epsilon_{cr}$ appears to be rather sensitive to the initial parameters of the ferroelectric state after quenching from the paraelectric one. Its dependence on the initial magnitude of the spatial polarization variations, characterized by the dispersion $D(0)$, and on the initial spatial scale of the polarization variations, characterized by the Gauss parameter $r_c$, is presented in Fig. 12. It is seen that the coercive field rises monotonically with the increase of both $D(0)$ and $r_c$.

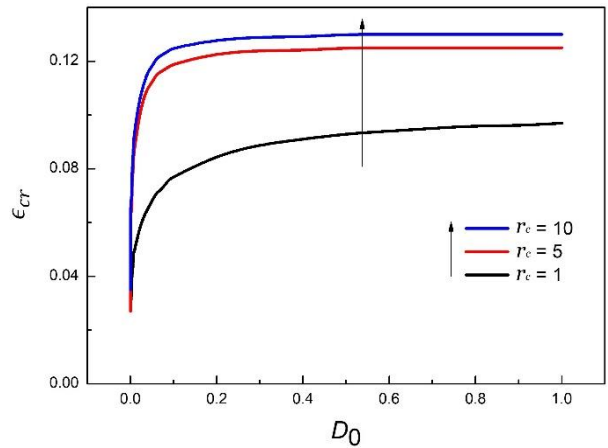

Fig. 12. The dependence of the coercive field on the initial value of the polarization variance $D(0)$ for different values of the initial characteristic length $r_c$.

## 5.2. Development from the initial correlation function of an exponential shape

We consider now the solution of the equation system (12) for the variables $\bar{\pi}(\tau)$ and $D(\tau)$ using the initial correlation function $K(s, 0)$ of an exponential shape from Eq. (23) reflected in the correlation length $L(\tau)$ given by Eq. (27). The initial conditions for Eqs. (12) are again $\bar{\pi}(0) = 0$ and $D(0) = 0.1$ while the other involved parameters are $\xi = 1, \eta = 10, \alpha_z = 0.625$. The evolution of the system with time $\tau$ as a parameter is presented on the phase diagram in coordinates $(\bar{\pi}, D)$ for different values of the field $\epsilon_a$ as indicated in Fig. 13.

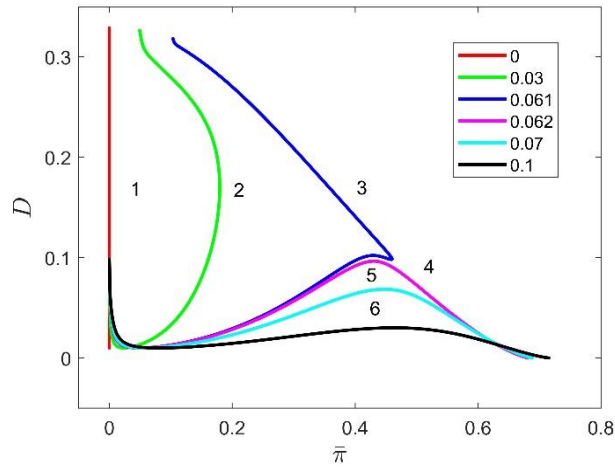

Fig. 13. Phase trajectories of the system starting from the initial values $\bar{\pi}(0) = 0$ and $D(0) = 0.1$ for the parameter values $\xi = 1, \eta = 10, \alpha_z = 0.625$. Curves 1–6 correspond to the values of the dimensionless external electric field in the sample $\epsilon_a = \{0; 0.03; 0.061; 0.062; 0.07; 0.1\}$, respectively.

The pattern of trajectories in Fig. 13 reminds of that in Fig. 9 for the initial Gaussian correlation function and demonstrates the same bifurcation physical behavior deciding between the single-domain and the multi-domain final state. The coercive field is somewhat smaller than in Fig. 9 and amounts to $\epsilon_{cr} = 0.0615$.

The explicit time dependences of $\bar{\pi}(\tau)$ and $D(\tau)$ are presented for different values of the applied field in Figs. 14 (a) and (b), respectively. The kinetics of the polarization structure formation represented by $\bar{\pi}(\tau)$ and $D(\tau)$ demonstrates, in principle, the same features as in the case of the initial Gaussian correlation function in Fig. 10.

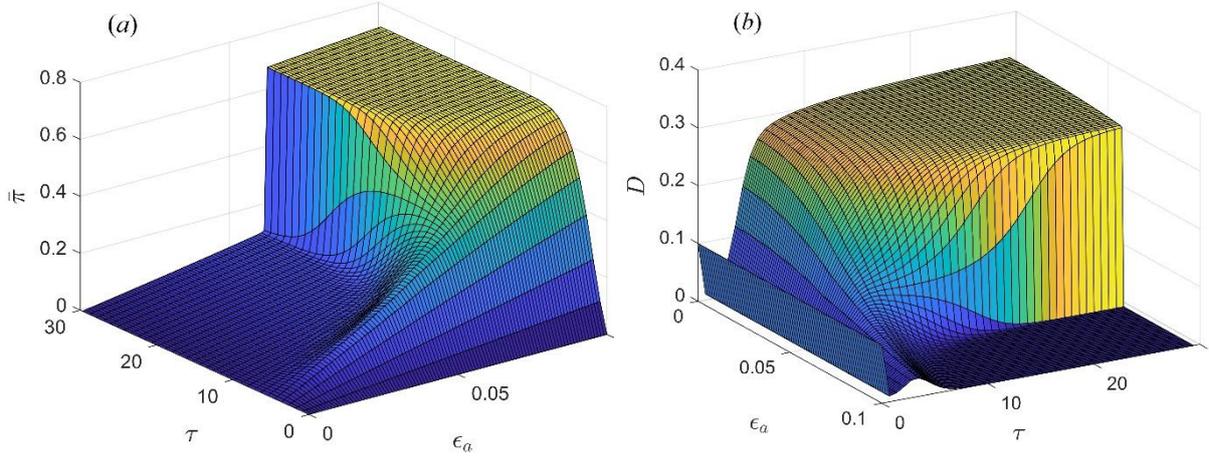

Fig. 14. (a) The time development of the mean polarization for different magnitudes of the electric field. (b) The time development of the polarization variance for different magnitudes of the electric field. All other parameters are the same as used in Fig. 13.

The magnitude of the initial disorder $D(0)$ plays again a significant role for the system evolution as is shown in Fig. 15. The phase diagram on the ($\bar{\pi},D$)-plane demonstrates the system development starting from different magnitudes of the initial disorder within the range $0.1 < D(0) < 0.9$ when the applied field $\epsilon_a = 0.08$ is fixed. It is seen again that variance $D$ in the incubation stage, to which the system falls after the start, increases monotonically with $D(0)$.

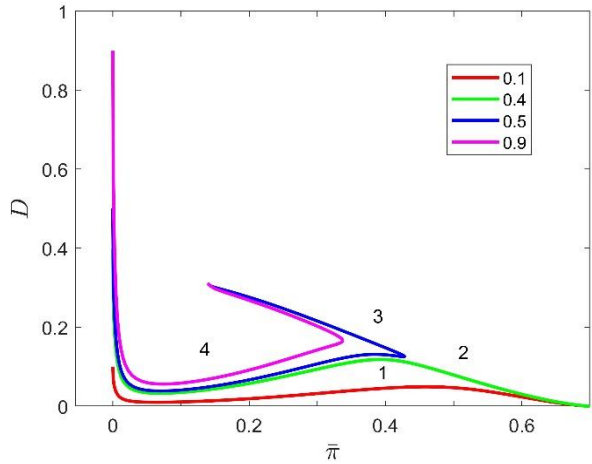

Fig. 15. Phase trajectories of the system starting from the initial value $\bar{\pi}(0) = 0$ under the applied field of $\epsilon_a = 0.08$ for the parameter values $r_c = 1, \eta = 10, \alpha_z = 0.625$. Curves 1–6 correspond to the values of the initial disorder $D(0) = \{0.1; 0.3; 0.4; 0.9\}$, respectively.

The coercive field $\epsilon_{cr}$ dependence on the initial parameters of the ferroelectric state after quenching from the paraelectric one exhibits the same trends as for the Gaussian initial state in Fig. 12. Its dependence on the initial magnitude of the spatial polarization variations $D(0)$ and on the initial spatial scale of the polarization variations, characterized by the parameter $\xi$, is presented in Fig. 16. Similar to Fig. 12, the coercive field rises monotonically with the increase of both $D(0)$ and $\xi$.

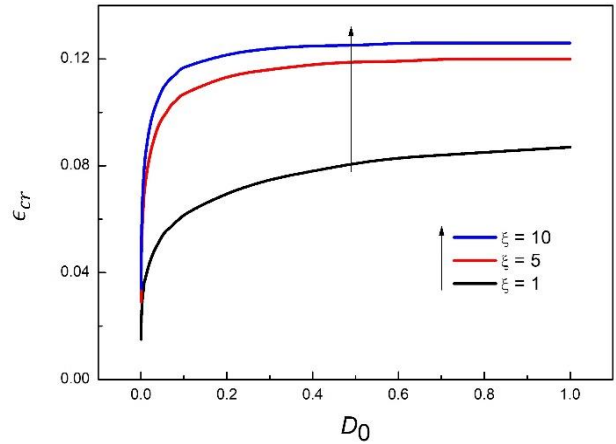

Fig. 16. The dependence of the coercive field on the initial value of the polarization variance $D(0)$ for different values of the characteristic length $\xi$.

### 5.3. Development from the initial correlation function of an error function type

Let us consider now the solution of the equation system (12) for the variables $\bar{\pi}(\tau)$ and $D(\tau)$ using the initial correlation function $K(\boldsymbol{s}, 0)$ of an error function type from Eq. (29) reflected in the correlation length $L(\tau)$ given by Eq. (33). The initial conditions for Eqs. (12) are again $\bar{\pi}(0) = 0$ and $D(0) = 0.1$ while the other involved parameters are $\xi = 1, \eta = 10, \alpha_z = 0.625$. The phase diagram in coordinates $(\bar{\pi}, D)$ of the system evolution for different values of the field $\epsilon_a$ is presented in Fig. 17.

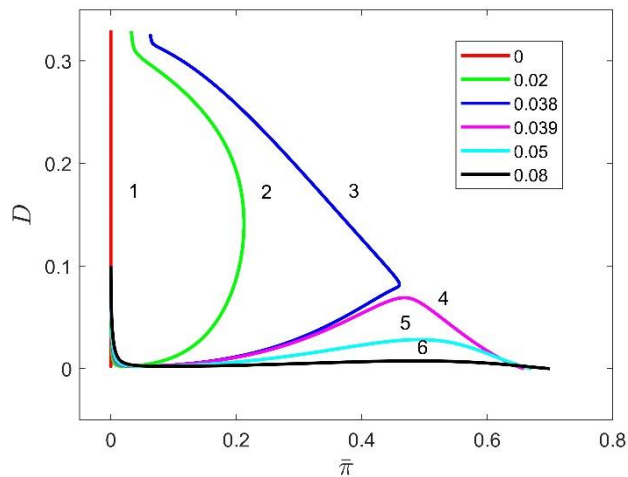

Fig. 17. Phase trajectories of the system starting from the initial values $\bar{\pi}(0) = 0$ and $D(0) = 0.1$ for the parameter values $\xi = 1, \eta = 10, \alpha_z = 0.625$. Curves 1–6 correspond

to the values of the dimensionless external electric field in the sample $\epsilon_a = \{0; 0.02; 0.038; 0.039; 0.05; 0.08\}$, respectively.

The phase diagram in Fig. 17 reminds of that in Figs. 9 and 13 and demonstrates the same bifurcation physical behavior deciding between the single-domain and the multi-domain final state. The coercive field is further reduced and amounts to $\epsilon_{cr} = 0.0385$.

The explicit time dependences of $\bar{\pi}(\tau)$ and $D(\tau)$ for different values of the applied field are shown in surface plots in Figs. 18 (a) and (b), respectively. The development of the polarization structure represented by $\bar{\pi}(\tau)$ and $D(\tau)$ exhibits, in principle, the same trends as in the cases of the initial Gaussian correlation function (Fig. 10) and exponential correlation function (Fig. 14), however, there are distinct quantitative features. Thus, starting from the initial value of $D(0) = 0.1$, the system falls down very fast to low values of $D \ll 1$.

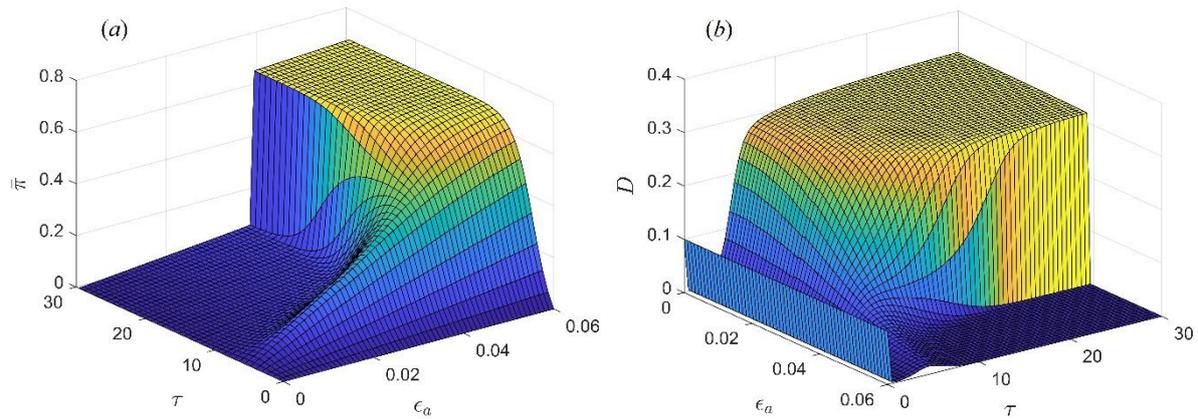

Fig. 18. (a) The time development of the mean polarization for different magnitudes of the electric field. (b) The time development of the polarization variance for different magnitudes of the electric field. All other parameters are the same as used in Fig. 17.

The magnitude of the initial disorder $D(0)$ strongly affects the system evolution as is shown in Fig. 19. The phase diagram on the ($\bar{\pi}$,$D$)-plane discloses the system development starting from different magnitudes of the initial disorder within the range $0.1 < D(0) < 0.9$ when the applied field $\epsilon_a = 0.05$ is fixed. It is seen that a very low variance $D$ in the incubation stage, to which the system falls after the start, increases monotonically with $D(0)$.

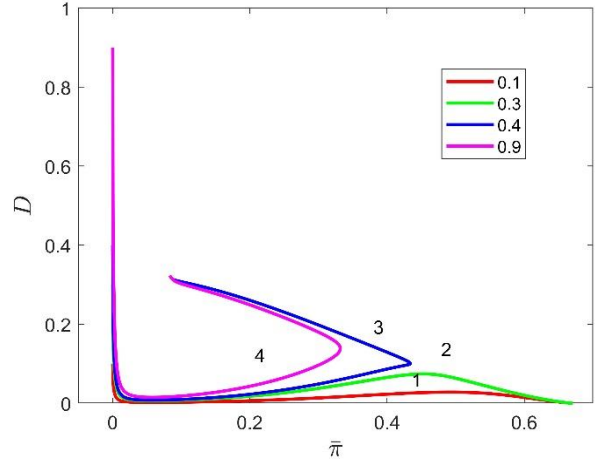

Fig. 19. Phase trajectories of the system starting from the initial value $\bar{\pi}(0) = 0$ under the applied field of $\epsilon_a = 0.05$ for the parameter values $r_c = 1, \eta = 10, \alpha_z = 0.625$. Curves 1–6 correspond to the values of the initial disorder $D(0) = \{0.1; 0.3; 0.4; 0.9\}$, respectively.

As well as in the previous cases of the initial correlation functions treated in sections 6.1 and 6.2, the coercive field $\epsilon_{cr}$ for the initial correlations of the error function type demonstrates strong dependence on the initial parameters of the ferroelectric state after quenching from the paraelectric one. The dependence on the initial magnitude of the spatial polarization variations $D(0)$ and on their initial spatial scale characterized by the length parameter $\xi$, is presented in Fig. 20. Similar to Figs. 12 and 16, the coercive field rises monotonically with the increase of both $D(0)$ and $\xi$ but comes slower to the saturation at higher values of $D(0)$.

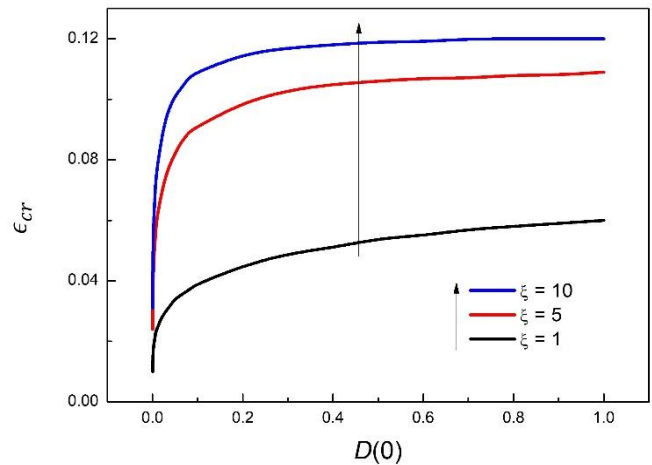

Fig. 20. The dependence of the coercive field on the initial value of the polarization variance $D(0)$ for different values of the characteristic length $\xi$.

### 6. Charge density correlations

As was established in recent experiments combined with phase-field simulations, the domain structures in uniaxial ferroelectrics PGO [3,33] and TGS [34] exhibit unexpected and exotic properties. Particularly, the head-to-head and tail-to-tail domain configurations, inevitable in these uniaxial systems, do not demonstrate the expected high polarization charges. This paradoxical feature was explained by the formation of opposing branched domains that reduce nominally charged domain walls to saddle points [33,34]. The polarization correlation functions evaluated above in section 4 do not directly reveal polarization configurations; however, the charge density correlation function, which can be derived from them, may indirectly characterize the charge state of the domain walls, as will be shown in the following.

The charge correlation function $M(\mathbf{s}, \tau) = \langle \rho_b(\mathbf{r} + \mathbf{s}, \tau) \rho_b(\mathbf{r}, \tau) \rangle$ can be easily derived from the polarization correlation function $K(\mathbf{s}, \tau)$ noting that the bound charge density $\rho_b = -\nabla \mathbf{P}$. Particularly, the spatial charge variance may be then evaluated as $\langle \rho_b(\mathbf{r}, \tau)^2 \rangle = M(0, \tau)$.

For the general case of an anisotropic multiaxial ferroelectric,

$$M(\mathbf{r}_1 - \mathbf{r}_2, \tau) = \langle \frac{\partial P_\alpha(\mathbf{r}_1, \tau)}{\partial r_{1,\alpha}} \frac{\partial P_\beta(\mathbf{r}_2, \tau)}{\partial r_{2,\beta}} \rangle = -\frac{\partial^2}{\partial r_{1,\alpha} \partial r_{1,\beta}} K_{\alpha\beta}(\mathbf{r}_1 - \mathbf{r}_2, \tau), \qquad (36)$$

where the correlation function $K_{\alpha\beta}(\mathbf{r}_1 - \mathbf{r}_2, \tau)$ denotes correlations between different polarization components. For the considered here uniaxial system, the charge correlation function in terms of the above introduced dimensionless units reads

$$M(\mathbf{s}, \tau) = -\frac{\partial^2}{\partial s_z^2} K(\mathbf{s}, \tau). \qquad (37)$$

Since the problem of the disappearing charges at the domain walls is hardly related to a specific initial shape of the polarization correlation function, the latter, in this section, will be taken in the Gauss form (18) which allows obtaining the results in a closed form. Thus, using the formula derived for the longitudinal correlation coefficient along the polarization axis, $C_\parallel(s_z, \tau) = K(s_z, \tau)/D(\tau)$ [21], the charge correlation function in the direction $\mathbf{s} = (0, s_z)$ results as

$$M_\parallel(s) = -\frac{D(\tau)\sqrt{2\eta\tau}}{\mathrm{erf}(\sqrt{2\eta\tau})} \left\{ \frac{2\eta\tau}{(L\zeta)^2} \left[ \frac{\mathrm{erf}(\sqrt{\zeta})}{2\sqrt{\zeta}} \left( \frac{5s^2}{6L^2\zeta} - 1 \right) + \frac{\exp(-\zeta)}{3\sqrt{\pi}} \left( 1 - \frac{s^2}{3L^2} - \frac{7s^2}{6L^2\zeta} \right) \right] \right.$$
$$\left. + \frac{\exp(-\zeta)}{3\sqrt{\pi}L^2\zeta} \left[ 2 - \frac{5s^2}{3L^2}\left(1 + \frac{1}{\zeta}\right) + \frac{s^4}{9L^4}\left(1 + \frac{2}{\zeta} - \frac{2}{\zeta^2}\right) \right] \right\} \qquad (38)$$

with the combined variables $L = \sqrt{(r_c^2 + 4\tau)/3}$ and $\zeta = 2\eta\tau + s^2/6L^2$, introduced for convenience.

In the plane transverse to the polarization direction, $\mathbf{s} = (\mathbf{s}_\perp, 0)$, the charge correlation coefficient can be derived in the integral form by substituting Eqs. (6) and (10) in Eq. (37),

$$M_\perp(\mathbf{s}_\perp, \tau) = \frac{3\sqrt{6}}{\pi} \frac{\sqrt{2\eta\tau}}{\text{erf}(\sqrt{2\eta\tau})} \frac{D(\tau)}{L^2} \int_0^\infty dk\, k^4 \exp\left(-\frac{3}{2}k^2\right) \int_{-1}^1 du\, u^4 \exp(-2\eta\tau u^2) J_0\left(\frac{k\sqrt{1-u^2}}{L} s_\perp\right). \tag{39}$$

Spatial dependences of the longitudinal and transverse charge correlation functions, Eq. (38) and Eq. (39), are shown for a range of time moments in Fig. 21(a) and (b), respectively. Since the variance $D(\tau)$ saturates fast with time (see Fig. 10), its value is approximated for all chosen times by $D(\tau) \cong 0.3$.

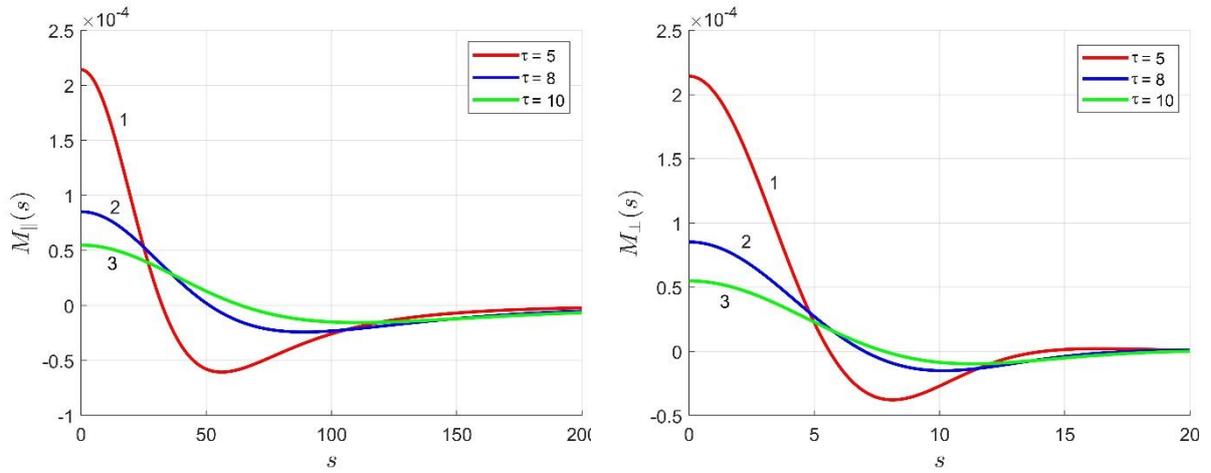

Fig. 21. (a) The spatial dependence of the longitudinal correlation function, Eq. (38), for the time moments $\tau = 5, 8$ and $10$ represented by the curves 1,2 and 3, respectively. (b) The spatial dependence of the transverse correlation function, Eq. (39), for the time moments $\tau = 5, 8$ and $10$ represented by the curves 1,2 and 3, respectively.

By determining the characteristic size of the domains from Fig. 21 at half the height of the correlation functions, as is usually done in the experiment [28-32], their longitudinal size at time $\tau = 10$ can be estimated as $L_\parallel \cong 40$ and their transverse size at the same time as $L_\perp \cong 5$. Let us imagine typical for TGS stripe domains of an approximately rectangle cross-section of the size $L_\perp \times L_\parallel$. The head-to-head and tail-to-tail domains generate at the domain wall the surface charge $\cong 2P_s$. Using the dimensionless units of the surface charge normalized to $P_s$, the density of the space charge due to the charged domain walls can be estimated as $2/L_\parallel \cong 5 \times 10^{-2}$. In contrast, the maximum value of the charge density variance observed in Fig. 21 at $\tau = 10$ amounts to $M(0, \tau) \cong 5 \times 10^{-5}$ corresponding to the characteristic charge density of $\sqrt{M(0,\tau)} \cong 7 \times 10^{-3}$. Since our model does not contain free charge carriers, this means that the bound charges have to be effectively reduced by one order of the magnitude by some structural organization of domains, which can be, for example, the saddle-point domains structures observed in Refs. [33,34].

## 7. Discussion

In this work, we have studied the effect of various initial conditions, formed after quenching from the high-temperature paraelectric to the low-temperature ferroelectric state, on the temporal evolution of the system and its physical properties. To this end we used the exactly solvable stochastic model of a uniaxial ferroelectric, which allows for calculation of the time dependent characteristic length and the two-point polarization correlation function [21] as well as all the other involved correlation functions between the polarization and the electric field components [27], assuming that they all are Gauss random variables. The model assumes domination of the quenched disorder over the thermal fluctuations, which are known to be quite relevant for the ferroelectric properties [35]. The thermal fluctuations, however, were shown to be negligible if the magnitude and the spatial scale of the quenched polarization disorder are high enough and temperature is not too close to the ferroelectric phase transition temperature [21]. Thus, the model allows investigation of the polarization formation kinetics in a wide temperature and the applied electric field range for different initial conditions in the material after quenching.

Temporal dependences of the characteristic correlation length $L(\tau)$ and the polarization correlation coefficient $C(s,\tau)$ of the domain structures developing after quenching were carefully investigated experimentally in uniaxial TGS ferroelectrics over thirty years [28-32]. Particularly, the scaling dependence at small distances $s$, $C(s,\tau) \cong 1 - s/2L(\tau)$, was established in all the works [28-32] but never explained theoretically. Since the previously used Gaussian form of the initial correlation function, Eq. (18), did not allow to describe the linear spatial behavior of $C(s,\tau)$ near the origin [21], we considered in the current paper alternative shapes of the initial correlation function of the exponential and the complimentary error function forms, which are linear at small $s$. For all considered shapes of the initial correlation function exact expressions for $L(\tau)$ and $C(s,\tau)$ were obtained in sections 4.1-4.3.

All the theoretical temporal dependences of $L(\tau)$ for different initial states satisfactorily described the experimental observations by Tomita et al. [15], as is shown in Figs. 2 and 5, with some advantage in favor of the exponential dependence, as can be seen in Fig. 2. The time-dependent correlation coefficients $C(s,\tau)$ derived for their different initial shapes in sections 4.2-4.3 roughly described the experimental data by Tomita et al. [15] including the properly located region of negative values and the proper trend of variation with time, as is seen in Figs. 4 and 7. However, contrary to the expectations, the initial linear spatial dependence near the origin quickly transformed with time to a form with a smooth maximum thus contradicting the experimental observations. Considering the transformation of the theoretical correlation functions with time in Figs. 3 and 6, this behavior seems to be a common feature determined by the structure of the governing equations (8).

Furthermore, a comparison of the correlation functions for the model stripe domain structures with zero and non-zero domain wall thicknesses in section 4.4 suggested, that the apparent linear spatial dependence of $C(s,\tau)$ might result from the way of experimental data treatment in the papers [28-32]. In all these works the regions of opposite polarizations were interpreted in the spirit of the Ising model by assigning the values of the scalar order parameter +1 or -1 thus assuming the zero thickness of the domain walls.

Another task of the current work was to establish the role of the initial conditions on the domain formation kinetics when subjected to the external electric field. The solutions (10-11) reduced the system of integrodifferential equations (8) for the correlation function and the mean polarization to the system of nonlinear differential equations (12) for the mean polarization $\bar{\pi}(\tau)$ and the polarization variance $D(\tau)$. The latter system was then studied numerically in section 5 for different initial correlation functions, different magnitudes of the initial disorder, characterized by the parameter $D(0)$, and its different initial spatial scales. The numerical analysis has revealed the following traits.

In all cases, starting from an initial value of $D(0)$, the system fell down to a much smaller value of the polarization variance, exhibiting a kind of homogenization, and then, after some incubation period, developed towards a single- or a multi-domain state depending on the magnitude of the applied electric field (Figs. 9-10, 13-14 and 17-18). The properties and the duration of the incubation period depending on the $D(0)$ value are observable in Figs. 11, 15 and 19. The coercive field, deciding between the single and multi-domain state, appeared to be dependent on the initial shape of the correlation function $K(s, 0)$ and in all cases exhibited a monotonic increase with both the disorder magnitude $D(0)$ and its spatial scale defined by the parameters $r_c$ or $\xi$ (see Figs. 12,16 and 20). Since the properties of the initial disordered state can be controlled to a certain extent experimentally by, say, cooling rate [1-5], this presents a tool for achieving desired ferroelectric properties. The typical values of the dimensionless coercive field $\epsilon_{cr}$ are in the region 0.05-0.1 which makes about 0.02-0.04 of the thermodynamic coercive field [36]. For a comparison, the experimental coercive field in TGS at room temperature is about 60 kV/m [37] making about 0.004 of the thermodynamic coercive field of

15 MV/m for TGS [36]. Thus, the current theory still overestimates the magnitude of the coercive field by one order of the magnitude, which is not surprising since it does not account for many possible realistic mechanisms of domain nucleation, for example, at the sample surface, at different defects etc.

Astounding is that the temporal development of the correlation length $L(\tau)$, Eqs. (15,22,27,33), turned out to be independent on both the applied electric field and the polarizability of the material $\eta$. It depends only on the characteristics of the initial state represented by the initial correlation function $K(s, 0)$ and its parameters $r_c$ or $\xi$. Particularly, it remains the same for the applied electric fields below and above the coercive field, meaning that the growing single-domain and multi-domain structures exhibit the same transient domain size. Furthermore, the correlation length exhibits the same asymptotic behavior $L(\tau) \sim \sqrt{4\tau/3}$ at $\tau \to \infty$ independently of the shape of the initial correlation function $K(s, 0)$. Apparently, this behavior is determined by the structure of equations (8) but not by the initial conditions. In contrast, the behavior of $L(\tau)$ at small times $\tau \ll 1$ is dependent on the initial state. For $K(s, 0)$ of the Gaussian type, $L(0) = r_c/\sqrt{3}$ and is finite. For the other two initial correlation function shapes, the correlation length vanishes at $\tau \to 0$ as $L(\tau) \sim (\xi^2\tau)^{1/4}$ and is dependent on the parameter $\xi$. The preferred choice of $K(s, 0)$, however, cannot be unambiguously justified by a comparison with experiments because the transient data were always reported after some time lag.

The correlation coefficient $C(s, \tau)$ also exhibited universal features being independent of the applied electric field which may seem strange. However, the correlation function $K(s, \tau) = C(s, \tau)D(\tau)$ strongly depends on the electric field together with the polarization variance $D(\tau)$, as is seen in Figs. 10, 14 and 18, and, thus, develops distinctly at electric fields above and below the coercive field.

As a by-product, the knowledge of the polarization correlation function allowed an insight into the enigmatic problem of charged domain walls in uniaxial ferroelectrics. Namely, the charge density correlations and fluctuations appeared to be substantially reduced, indicating the charge on the domain walls one order of the magnitude lower than expected for the head-to-head and tail-to tail domain walls in uniaxial ferroelectrics.

## 8. Conclusions

Based on the exactly solvable stochastic model of the domain structure formation in uniaxial ferroelectrics dominated by the quenched polarization disorder [21], we studied the kinetics of the domain development from different initial disordered states characterized by distinct polarization correlation functions, polarization fluctuation magnitudes and characteristic spatial scales. For all cases, analytical results were obtained for the time-dependent characteristic length of the domain structure and the two-point polarization correlation coefficient, which are in satisfactory agreement with experiments. These two temporal characteristics of the stochastic system appeared to be quite universal being independent on the applied electric field. On the other hand, the polarization correlation function, proportional to the polarization variance, is dependent on the electric field and exhibits distinct behavior below and above the coercive field deciding between the multi-domain and the single-domain final state of the system. The coercive field, in turn, can be controlled by the parameters of the initial quenched state and increases with the increasing polarization fluctuation magnitude and spatial scale of fluctuations.

## Acknowledgements
O.Y.M. is grateful for the financial support by the Marie Sklodowska-Curie Actions for Ukraine (№ 1233427). This work was supported by the Deutsche Forschungsgemeinschaft (German Research Society, DFG) via Grant No. 405631895 (GE-1171/8-1).